\documentclass[prl,aps,groupedaddress,showpacs,]{revtex4}
\usepackage{amssymb,amsmath,amsfonts,epsfig,graphicx,latexsym,bm,color,subfigure}

\begin{document}
\title{Pulse delay time statistics in a superradiant laser with calcium atoms}
\author{Torben Laske, Hannes Winter, and Andreas Hemmerich\footnote{e-mail: hemmerich@physnet.uni-hamburg.de} }
\affiliation{Institut f\"ur Laser-Physik and Zentrum f\"ur Optische Quantentechnologien, Universit\"at Hamburg, D-22761 Hamburg, Germany}
\date{\today}

\begin{abstract}
Cold samples of calcium atoms are prepared in the metastable $^{3}$P$_{1}$ state inside an optical cavity resonant with the narrow band (375 Hz) $^{1}$S$_{0} \rightarrow ^{3}$P$_{1}$ intercombination line at 657 nm. We observe superradiant emission of hyperbolic secant shaped pulses into the cavity with an intensity proportional to the square of the atom number, a duration much shorter than the natural lifetime of the $^{3}$P$_{1}$ state, and a delay time fluctuating from shot to shot in excellent agreement with theoretical predictions. Our incoherent pumping scheme to produce inversion on the $^{1}$S$_{0} \rightarrow ^{3}$P$_{1}$ transition should be extendable to allow for continuous wave laser operation.
\end{abstract}

\bibliographystyle{prsty}

\maketitle
Conventional lasers typically operate in the so-called \textit{good cavity} limit, where the resonance bandwidth of the feedback cavity is by far more narrow than the spectral width of the gain profile. The achievable emission bandwidth is presently approaching fundamental limitations by intrinsic thermal fluctuations of the cavity materials \cite{You:99, Kes:12, Mat:17}, which is one of the obstacles for further improvements of the precision of atomic clocks \cite{Lud:15}. An alternative approach to circumvent these limitations relies on the use of an ultra-narrow bandwidth gain material, as provided by two-electron atoms like calcium or strontium, in combination with a comparatively large cavity bandwidth. In this so-called \textit{bad cavity} regime, the average intra-cavity photon number can be kept small and even well below unity such that the intra-cavity field cannot establish coherence, as in the good cavity regime. Here, it is rather the long-lived atomic polarization providing the phase memory necessary to form coherence by superradiant emission, with the result of a sensitivity to technical noise sources reduced by many orders of magnitude. Bad cavity lasers, also referred to as superradiant lasers, are a subject of ongoing theoretical \cite{Lax:66, Hak:84, Haa:93, Mei:09} and experimental \cite{Kup:94, Boh:12} research. In the recent past, superradiant lasing has undergone a renaissance in connection with the use of ultranarrow band intercombination lines of alkaline-earth atoms \cite{Che:05, Che:09, Mei:09, Nor:16a, Nor:16b}, which could provide extremely low emission bandwidths in the sub-millihertz regime.  

Superradiant emission of an inverted system in free space has been studied since the fifties \cite{Dic:54, Ern:68, Reh:71, Bon:71, Haa:72, Bon:75, Gla:75, Gro:82} followed by first observations in the optical domain in the seventies \cite{Skri:73, Gro:76, Cah:79}. More recently, a new line of research has been concerned with the collective light scattering by dense ultra-cold samples of atoms in free space as well as inside optical cavities \cite{Ino:99,Sla:07, Kes:14, Mul:16}. On a macroscopic level, a completely inverted system represents an unstable equilibrium. Its decay is triggered by microscopic quantum fluctuations, which translate into macroscopic shot to shot delay time fluctuations of classical superradiant light pulses. This phenomenon has been theoretically studied \cite{Haa:81, Gro:82} but a quantitative comparison with experiments is yet missing. In this article we report the first pulsed superradiant laser with bosonic calcium ($^{40}$Ca) atoms. This is achieved by providing inversion with respect to the narrow band ($\Gamma/2 \pi = 375\,$Hz) $^{1}$S$_{0} \rightarrow ^{3}$P$_{1}$ intercombination line at $657\,$nm \cite{Rie:04}. Hyperbolic secant shaped pulses are observed with a temporal delay that fluctuates from shot to shot, thus reflecting the initial quantum stage of the pulse evolution. The stochastic nature of superradiance is studied quantitatively by measuring the pulse delay time statistics for different numbers of participating emitters. We find excellent agreement with an analytical model that does not require the adjustment of fitting parameters. In contrast to a recent first experimental realization of superradiant lasing with strontium atoms \cite{Nor:16a, Nor:16b}, we use an incoherent pump process to provide inversion, which should allow an extension to continuous wave operation.

\begin{figure}
\includegraphics[scale=0.4, angle=0, origin=c]{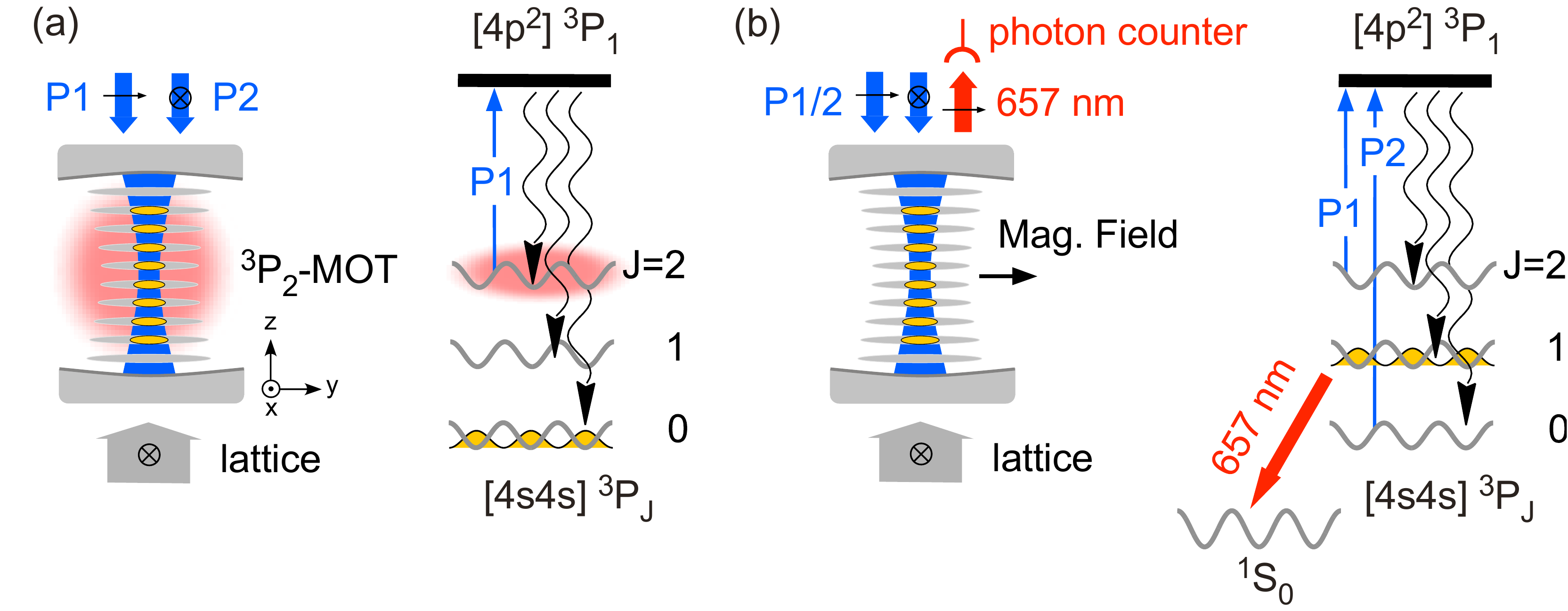}
\caption{\label{Fig.1} The two steps for preparation of inversion are sketched in (a) and (b). An intracavity lattice at 800.8 nm is formed with linear polarization along the $x$-axis. P1 and P2 denote optical pumping lasers with linear polarizations along the $y$-axis and $x$-axis, operating at 432 nm and 429 nm, respectively. Laser emission at $657\,$nm is recorded with a photon counter.}
\label{fig:prep}
\end{figure}

The preparation of inversion on the $^{1}$S$_{0} \rightarrow ^{3}$P$_{1}$ transition at $657\,$nm proceeds in two steps illustrated in Fig.~\ref{fig:prep}: Initially, a magneto-optic trap ($^{3}$P$_{2}$-MOT), using the 57~kHz $^{3}$P$_{2} \rightarrow ^{3}$D$_{3}$ closed cycle transition at 1978~nm, prepares about $2 \times 10^{8}$ atoms at a temperature of $200\,\mu$K in the $^{3}$P$_{2}$ state. Details are found in Refs.~\cite{Gru:01, Gru:02, Han:03, Han:06}. As sketched in Fig.~\ref{fig:prep}(a), the cold cloud of atoms produced in the $^{3}$P$_{2}$-MOT is superimposed upon a longitudinal mode of a linear cavity with a finesse $F=2200$, a free spectral range of 2.5~GHz, a transmission resonance bandwidth $\kappa / \pi = 2260\,$kHz, and a ($1/e^{2}$) waist $w_0 = 190\,\mu$m (for light at a wavelength of 657 nm). The cavity exhibits a Purcell factor $\eta \equiv \frac{24 F}{\pi k^2 w_0^2} = 0.0051$ with $k= 2\pi/657\,$nm \cite{Pur:46}. A laser beam at 800.8~nm, i.e., the magic wavelength \cite{Kat:03} for the $^{1}$S$_{0} \rightarrow ^{3}$P$_{1}, m=0$ transition for $\sigma_{\pm}$ polarized light, is coupled to the cavity to create an intra-cavity optical lattice potential. A laser beam (P1) at 432~nm, aligned with the cavity mode with a waist of $w_0 = 100\,\mu$m, pumps atoms within a small tube around the lattice axis into the $^{3}$P$_{0}$ state, where they are trapped in the lattice potential. The intensity of this beam is low such that only atoms passing the pump volume at low velocities and hence interacting with this beam sufficiently long, are pumped \cite{Yan:07, Hal:12, Hal:13}. Thus, up to several $10^5$ atoms in the $^{3}$P$_{0}$ state are typically trapped in the lattice potential with a temperature of about $100\,\mu$K and a peak density of $10^{10}$cm$^{-3}$. Adjustment and technical fluctuations of the atom number are discussed in more detail in Ref.~\cite{Sup}. The trapped atom cloud extends over $2.5\,$mm along the lattice axis corresponding to several thousand lattice sites with an average population of a single pancake-shaped site of several ten atoms. Hence, contact interaction between atoms may be completely neglected. Also, in contrast to superradiant scattering in Bose-Einstein condensates, at these low densities, no complex light propagation dynamics occurs \cite{Den:10}.

In a second phase of the preparation protocol, sketched in Fig.~\ref{fig:prep}(b), the $^{3}$P$_{2}$-MOT is switched off, a homogeneous magnetic field of a few Gauss is oriented along the $y$-axis and an additional laser at 429~nm (P2), also adjusted along the cavity axis, however linearly polarized in $x$-direction, is activated during $50\,\mu$s, in order to rapidly pump all $^{3}$P$_{0}$ atoms into the $^{3}$P$_{1}$ state. According to a rate equation model of the optical pumping dynamics, only about $25 \%$ of the $^{3}$P$_{1}$ atoms are transferred into the relevant $m=0$ state, which acts as the upper laser level. The applied homogeneous magnetic field shifts the magnetic $m=\pm 1$ atoms several MHz out of resonance with the cavity. The lattice potential provides the same light-shift for the $^{3}$P$_{1}, m=0$ and $^{1}$S$_{0}$ states \cite{Deg:04} and operates well within the Lamb-Dicke regime, such that the Doppler effect is suppressed along the $z$-direction to first order. Hence, if the cavity is tuned into resonance, superradiant emission of a large fraction of the $^{3}$P$_{1}, m=0$ atoms can arise. Precisely controlled tuning of the laser cavity resonance with respect to the $^{1}$S$_{0} \rightarrow ^{3}$P$_{1}$ transition frequency is achieved by actively stabilizing the cavity to a diode laser beam at $780.2\,$nm locked to a Doppler-free resonance of rubidium atoms and sent through a electro-optic fiber modulator (EOFM) tunable between $400$ and $1000\,$MHz. Adjusting the EOFM driving frequency, the cavity resonance is adjusted to match the atomic resonance at $657\,$nm (See Ref.~\cite{Sup} for details). The lattice frequency is actively stabilized to the longitudinal mode of the laser cavity that is closest to the magic wavelength, i.e., less than half of the free spectral range, which amounts to $1.25\,$GHz.

\begin{figure}
\includegraphics[scale=0.3, angle=0, origin=c]{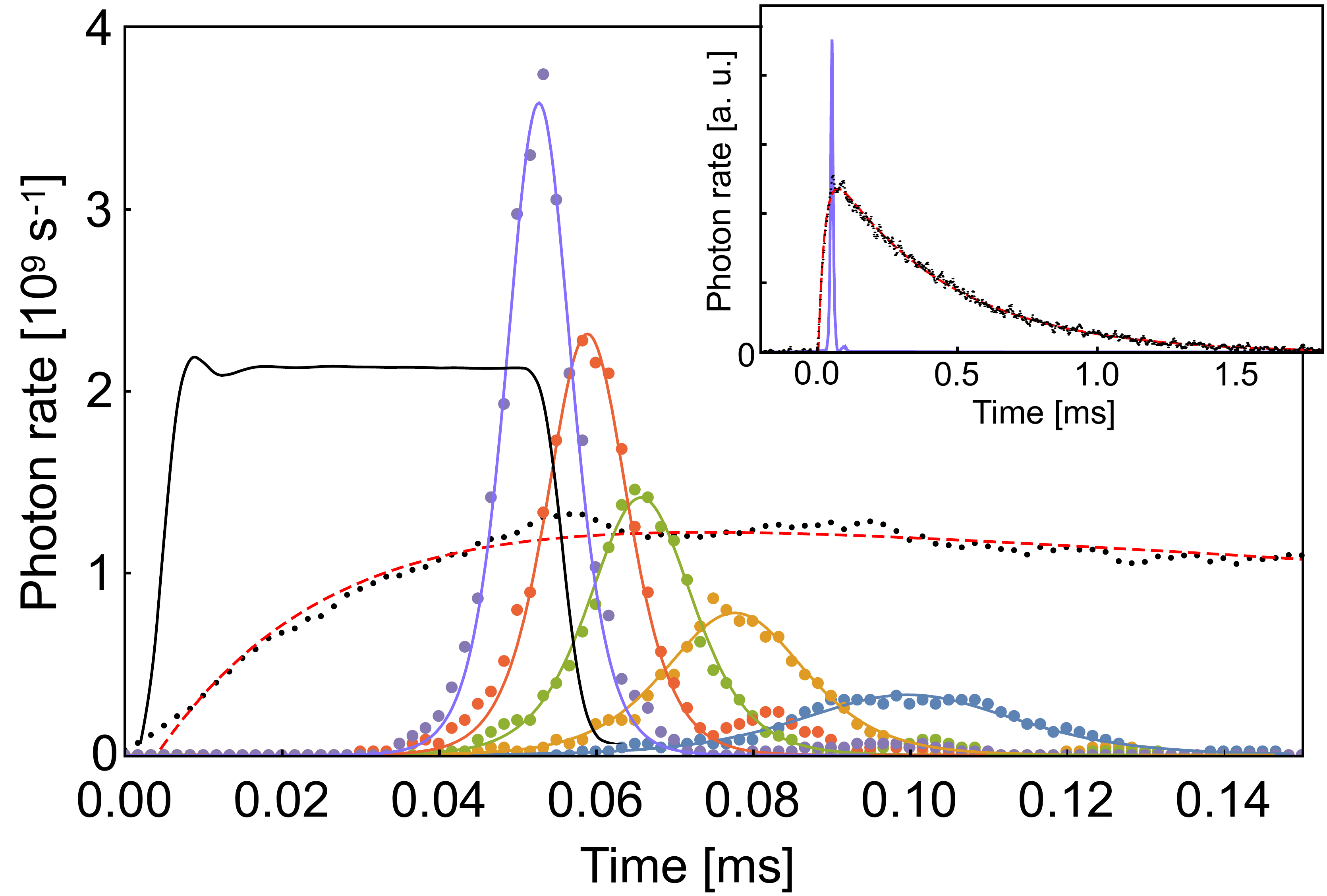}
\caption{\label{Fig.2} The inset (upper right corner) compares the natural non-cooperative exponential decay (black dots) with the case when a short ($ \approx  10\,\mu$s) superradiant pulse is emitted (blue graph). The red dashed line is a fit with two exponential functions as explained in the text. In order to present both graphs in the same plot, their vertical axes are scaled differently. The main panel shows superradiant light pulses for atom numbers $N_0 = 12800, 19700, 26500, 34000, 42300$ from right to left. The solid black line indicates the pump pulse that acts to populate the $^{3}$P$_{1}$ state. The black dots, modeled by the dashed red line graph, repeat the natural decay curve of the inset.}
\label{fig:pulse}
\end{figure}

After preparing an inverted sample of several $10^4$ atoms in the metastable $^{3}$P$_{1}, m=0$ state, we observe a superradiant pulsed emission (linearly polarized along the $y$-axis), which brings a significant fraction of the atoms into the  $^{1}$S$_{0}$ ground state in a time much shorter than the natural life time of the $^{3}$P$_{1}$ state. The inset in the upper right corner of Fig.~\ref{fig:pulse} compares the natural non-cooperative exponential decay with an observed life time of $420\,\mu$s (black dots approximated by red dashed line) with the case when a short ($ \approx  10\,\mu$s) superradiant pulse is emitted (blue graph). In both cases a sample of metastable calcium atoms in the $^{3}$P$_{1}$ state is prepared with the cavity tuned into resonance with the $^{1}$S$_{0} \rightarrow ^{3}$P$_{1}$ transition only in the latter case. The non-cooperative emission into free space is observed in the $xy$-plane at an angle of $22.5^\circ$ with respect to the $y$-axis. The main panel of Fig.~\ref{fig:pulse} zooms in to highlight the first $150\,\mu$s showing superradiant light pulses with five different peak photon numbers and peak times. Each trace represents a single-shot implementation. The observed pulses can be well fitted with hyperbolic secants derived from a semi-classical analytical model outlined below and in more detail in Ref.~\cite{Sup}. Two parameters are determined from these fits: the number of collectively emitting atoms $N_0$ and the time $t_p$ when the pulse attains its maximum. For the shown pulses, $N_0 = 12800, 19700, 26500, 34000, 42300$ from right to left. Besides $N_0$ and $t_p$ the hyperbolic secant fit model comprises the Purcell factor of the cavity $\eta$, the natural linewidth $\Gamma$ of the used transition and the bunching parameter $B$. The latter accounts for a reduction of the atom-cavity coupling strength resulting from the fact that the period of the atomic grating held by the magic lattice (800.8~nm) and that of the intra-cavity standing wave at the emission wavelength (657.5~nm) are not commensurate. For a homogeneous atomic distribution the value of $B$ should be 1/2, while our experimental data are best  described by $B \approx 0.65$. This value, which matches with a more elaborate analysis in Ref.~\cite{Hu:15}, is used in all fits. Taking into account the losses between the cavity and the photon counter and the detector efficiency, the specified counting rate is calibrated to indicate the total rate of photons leaving the cavity through both mirrors. A counting rate of $10^{9}\,$s$^{-1}$ corresponds to an intra-cavity photon number $n \approx 70$.

\begin{figure}
\includegraphics[scale=0.40, angle=0, origin=c]{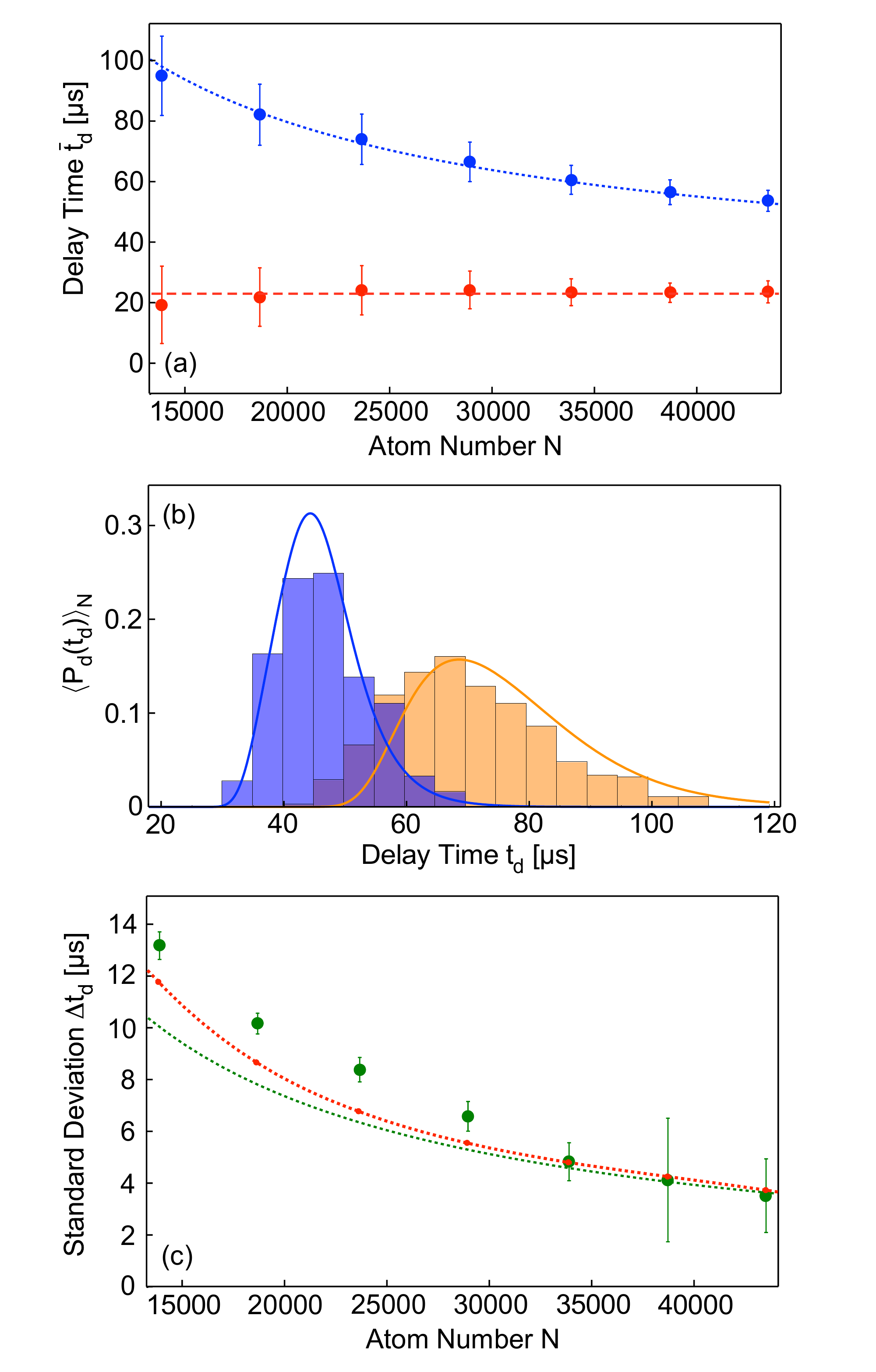}
\caption{\label{Fig.3} (a) pulse delay time versus $N_0$. The blue data points denote the observed pulse peak times $\bar{t}_d$. The red points arise, if the analytical model $\bar{t}_d$ is subtracted from the blue points leaving a small constant rest $t_0$. The blue dotted line graph shows $\bar{t}_d + t_0$. (b) Observed (histograms) and theoretically predicted (solid line graphs) delay time probability distributions (plotted versus $t_d$, i.e., with $t_0$ subtracted) for pulses with atom numbers around $N_1 = 3\times 10^4$ (blue) and $N_2 =1.5\times 10^4$ (yellow). (c) The green disks show the standard deviations of the observed pulse delay time versus the atom number $N_0$ for a partition dividing the $N_0$-axis into seven sectors. The error bars show the corresponding standard deviations of the mean. The dashed green line shows the theoretical expectation according to $P_d(t_d)$. The dashed red line in addition includes a systematic contribution due to the use of a finite partition.}
\label{fig:delay}
\end{figure}

A notable observation is that the average pulse peak time $\bar{t}_p$ grows with decreasing $N_0$, while for fixed $N_0$ the individually measured values of $t_p$ fluctuate around $\bar{t}_p$. In order to determine $\bar{t}_p$ and its dependence on the atom number $N_0$, we recorded several thousand pulses and plotted the averages over the observed pulse peak times $\bar{t}_p$ versus $N_0$ using a binning $N_0 \pm 2500$ for the atom number, such that each data point in the blue trace in Fig.~\ref{fig:delay}(a) represents several hundred pulses with nearly equal atom number. The error bars depict the standard deviations $\Delta t_p$, which, similarly as $\bar{t}_p$ itself, are observed to decrease for increasing $N_0$. In the red trace of (a), we subtract the analytical model $\bar{t}_d = (N_0 \eta B \Gamma)^{-1} \log(N_0)$, discussed below and in Ref.~\cite{Sup}. This model assumes instantaneous formation of complete inversion at some initial time $t_0$. For the choice $B = 0.65$, we remain with a practically constant temporal offset of $24\,\mu$s (red dashed line), which represents the time $t_0$ when inversion is effectively formed by the pump pulse. The blue dotted line graph shows $\bar{t}_d + t_0$. Apart from the choice of $B$, this procedure does not involve any parameter adjustment, thus confirming the validity of the theoretical description. 

In our experiment, according to the scheme illustrated in Fig.~\ref{fig:prep}(b), the upper laser level $^{3}$P$_{1}, m=0$ is in fact not instantaneously pumped, but rather loaded at a rate $R(t) \equiv  \frac{N_0}{\tau_p} e^{-t/\tau_p}$, where $\tau_p$ denotes the $1/e$ time for this process (see Fig.~\ref{fig:pulse}). The corresponding population of the upper laser level is $N(t) = \int_{0}^{t}ds R(s) = N_0 (1-e^{-t/\tau_p})$. The limited intensity available for the lasers P1 and P2 in Fig.~\ref{fig:prep}(b) gives rise to $\tau_p \approx 21\,\mu$s. An analysis deferred to Ref.~\cite{Sup} shows that the pulse peak time $\bar{t}_p$ can be nevertheless written as a sum $\bar{t}_p = t_0 + \bar{t}_d$, where $t_0$ only depends on $\tau_p$ but not on $N_0$, and $\bar{t}_d$ denotes the mean pulse delay time found for a scenario of instantaneous pumping, i.e., with $R(t) = 0$ and the upper laser level initially populated by $N_0$ atoms at time $t_0$. 

One may go beyond the determination of $\bar{t}_p$ and $\Delta t_p$ and consider the full pulse delay time probability distribution. Restricting ourselves to two different atom numbers $N_1 = 3\times 10^4$ and $N_2 =1.5\times 10^4$ we proceed as follows: The time axis is partitioned into time windows of $5\,\mu$s width and the number of pulses with atom numbers in the interval $\mathcal{N}_i \equiv [N_i -\delta N,N_i +\delta N ], i\in\{1,2\}$ with $\delta N = 5\times 10^3$ falling into each time window is counted. The histograms (normalized to unity) thus obtained for the two choices $N_i, i\in\{1,2\}$ are plotted in Fig.~\ref{fig:delay}(b). Our theoretical model predicts a delay time distribution $P_d(t_d) \equiv  N_0^2 B \eta \Gamma e^{-N_0 B \eta \Gamma t_d} \exp(-N_0 \, e^{-N_0 B \eta \Gamma t_d})$. The solid line graphs in Fig.~\ref{fig:delay}(b) show $\langle P_d(t_d) \rangle_{N_0 \in \mathcal{N}_i}$ where the bracket denotes averaging over all values of $N_0$ within the interval $\mathcal{N}_i$. Very good agreement between the observations and our analytical model is found, which does not rely on the adjustment of fit parameters. Finally, in Fig.~\ref{fig:delay}(c), the observed standard deviations of the pulse delay times as given by the error bars in (a) are plotted versus the atom number $N_0$ (green disks) and compared to the theoretical expectations according to $P_d(t_d)$ (dashed green line graph). Again, no parameter adjustment is applied. The observed fluctuations exceed the theoretical predictions for small atom numbers. A smaller part of this disagreement can be attributed to the choice of a relatively coarse partition in Fig.~\ref{fig:delay}(c), used to insure that each atom number class comprises a large number of pulses. The dashed red line graph shows the theoretically expected standard deviations including this partition-dependent contribution. See Ref.~\cite{Sup} for details. The remaining discrepancy between theory and experiment for small atom numbers might be due to the correspondingly large pulse delay times, which should increase possible contributions from technical fluctuations. Note also that we do not observe pulses with $N_0 < 10^4$, which indicates the presence of a possible threshold that could be attributed to inhomogeneous line broadening (see Ref.~\cite{Sup}). 

We now briefly summarize the theoretical model used to determine the delay time probability distribution $P_d(t_d)$ and the mean delay time $\bar{t}_d$ for the case of instantaneous formation of inversion. The extension to the case of non-instantaneous pumping is deferred to Ref.~\cite{Sup}. Our starting point is a modified semiclassical laser equation for $N_0$ two-level atoms all occupying the excited state at time $t=0$ (as for example in Eq.(8.8) on page 230 of Ref.~\cite{Mes:04}). We account for the fact that for a sample of $N_0$ atoms at positions $z_a$ along the axis of the lasing cavity mode, which is associated with an intensity distribution $\cos^2(k z_a)$, the effective coupling strength is reduced by the bunching parameter $B \equiv \frac{1}{N_0} \sum_a \cos^2(k z_a)$. For $0<t \ll \Gamma^{-1}$, a solution for the intra-cavity photon number is given by the pulse $n(t) = \frac{\eta B \Gamma}{8 \kappa} N_0^2 \textrm{sech}{^2}(\frac{1}{2} N_0 \eta B \Gamma(t-t_d))$ with $\textrm{sech}(z)$ denoting the hyperbolic secant. Setting $t=0$ and $n_0 \equiv n(0)$  leads to $n_0 \approx \frac{\eta B \Gamma}{2 \kappa} N_0^2 e^{-N_0 \eta B \Gamma t_d}$ and hence $d n_0 = -N_0 \eta B \Gamma n_0\,d t_d$. The delay time $t_d$ is determined by the initial photon number at $t=0$, which can be evaluated as follows. The mean rate of spontaneous photons released into the cavity at times much shorter than $(N_0 \eta \Gamma)^{-1}$ is $R_{\textrm{sp}} = N_0 \eta B \Gamma$, which is compensated by the photon loss rate $2 \kappa$ to yield the steady state photon number $\bar{n}_0 = \frac{R_{\textrm{sp}}}{2 \kappa}$. The rate $R_{\textrm{sp}}$ corresponds to an average time between successive spontaneous photons $\Delta t_{\textrm{sp}} = (N_0 \eta B \Gamma)^{-1}$, which is on the order of $10\,\mu$s. Assuming that $\bar{n}_0$ equals the initial photon number $n_0$ at $t=0$, we may evaluate the mean delay time as $\bar{t}_d = (N_0 \eta B \Gamma)^{-1} \log(N_0)$. We may go one step further and calculate the delay time distribution. Note that $\bar{n}_0$ is on the order of $10^{-2}$ and hence much smaller than unity. For this case the probability for $N_0$ atoms to spontaneously emit $n_0$ photons into the cavity, generally given by a binomial distribution, can be well approximated by an exponential. Hence, the probability of initially finding $n_0$ photons in the cavity is given by the normalized distribution $P_i(n_0) \equiv \frac{1}{\bar{n}_0} \exp(n_0/\bar{n}_0)$, which reproduces the previously determined steady state photon number $\bar{n}_0$. With the expressions of $t_d$ and $\frac{d t_d}{d n_0}$ as functions of $n_0$, discussed above, one finds $P_i(n_0) dn_0 = P_d(t_d) dt_d$, where $P_d(t_d)$ is the delay time distribution used in Fig.~\ref{fig:delay}(b). It is worthwhile to note that the expressions obtained for $P_d(t_d)$ and $\bar{t}_d$ are in accordance with reduced expressions, merely depending on $N_0$ and $\Gamma$, reported in Ref.~\cite{Gro:82} to model superradiance for a homogeneous atomic sample in free space with no cavity present. 

In contrast to Ref.~\cite{Nor:16b}, an incoherent pump process is used in our work to provide inversion, which should allow for an extension to continuous wave operation. Two different strategies into this direction could be followed. One option could be to use 657~nm radiation to re-excite ground state atoms to $^{3}$P$_{1}, m = \pm1$ and subsequently optically pump them with the help of 430~nm radiation to $^{3}$P$_{1}, m = 0$. A second option is to enable the loading of the magic lattice with $^{3}$P$_{2}$-atoms directly from the $^{3}$P$_{2}$-MOT. At present, this is not possible, because the magic lattice wavelength leads to a large negative light shift of the upper MOT level $^{3}$D$_{3}$, such that the MOT frequency is tuned to the blue side of the atomic resonance and hence lattice loading is impeded. This could be counteracted by an additional laser that selectively provides a compensating positive light shift to the $^{3}$D$_{3}$ state. Similar results as obtained in the present work should also be possible for the $^{1}$S$_{0} \rightarrow ^{3}$P$_{0}$ transition. This transition can be endowed with an finite linewidth in the subhertz range adjusted by adding a small admixture of the $^{3}$P$_{1}$ state to $^{3}$P$_{0}$ by means of a homogeneous magnetic field of a few Gauss \cite{Tai:06}. This could give rise to a superradiant laser with an adjustable extremely narrow bandwidth. Note that in this case the extension to continuous operation is straightforward, since the $^{3}$P$_{0}$ state can be continuously loaded from the $^{3}$P$_{2}$-MOT, as already shown in this work. 

\begin{acknowledgments}
This work was partially supported by DFG-He2334/15.1. We thank Claus Zimmermann for useful discussions.
\end{acknowledgments}

\section{Supplemental Material}
This supplemental material details the theoretical modeling of the observations discussed in the main text. We begin with briefly indicating our derivation of a semiclassical laser equation from Maxwell-Bloch equations. We account for the fact that the atoms are not necessarily located at the antinodes of the lasing cavity mode by introducing a bunching factor to describe the reduction of the coupling strength. For the case of instantaneous formation of complete inversion, the relevant hyperbolic secant solution is specified, which describes the emission of superradiant pulses. We model the initial quantum stage of the superradiant pulse evolution by the simple assumption that $N$ atoms spontaneously decay following a binomial probability distribution, which for sufficiently short times is well approximated by an exponential. This lets us determine the pulse delay time distribution used to interpret the experimental data shown in Fig.2 and Fig.3 of the main text. Finally, we consider the case, when the formation of inversion has a finite duration. We show that the dependence of the pulse delay time upon the particle number merely acquires an additional constant offset.  A short section on the prospects of continuous wave operation follows, before we close the discussion with a number of sections to detail technical issues and data analysis protocols.

\subsection{Semiclassical laser equation}
We consider two-level atoms confined in a quasi one-dimensional optical lattice, extending along the optical axis of a standing wave cavity. We begin with a description of the relevant semiclassical laser equation, accounting for the fact that the spatial distribution of the atoms in the laser mode yields a reduction of the coupling strength. Most of what is summarized here for our specific system can be found in standard textbooks (e.g. Refs.~\cite{Hak:84, Mes:04}). Apart from introducing a bunching factor $B$, to account for the reduced coupling strength, this section illustrates our definitions of the relevant quantities $\alpha_0 =  \textit{field per photon}$, $g =  \textit{Rabi-frequency per photon}$, $C = \textit{cooperativity}$, $\eta = \textit{Purcell factor}$, $V = \textit{mode volume}$, for which unfortunately no generally accepted conventions exists in the literature. 

Assuming a spatially and temporally constant light polarization we begin with the Maxwell-Bloch equations for a scalar real-valued electric field 
\begin{eqnarray}
\label{eq:MaxwellBloch0}
c^2 \Delta E(r,t) - \ddot{E}(r,t) = \kappa \dot{E} +\frac{1}{\varepsilon_0} \ddot{P}(r,t)
\end{eqnarray}
with a field decay term scaling with the rate $\kappa$ and a source term resulting from the real-valued polarization density $P(r,t)$. The real electric field and the polarization density are written in terms of complex quantities according to $E(r,t) = \frac{1}{\sqrt{2}} (\tilde{E}(r,t) e^{-i \omega_p t} + cc)$ and analogously for $P(r,t)$ with "cc" denoting the complex conjugate. It is assumed that the time dependence in $\tilde{E}(r,t)$ and $\tilde{P}(r,t)$ is by far slower than $\omega_p^{-1}$ such that one may write the approximate equation
\begin{eqnarray}
\label{eq:MaxwellBloch1}
\dot{\tilde{E}}(r,t) - \frac{i}{2} \left( \frac{c^2}{ \omega_p} \Delta + \omega_p + i \kappa \right) \tilde{E}(r,t)  = i \frac{\omega_p}{2 \varepsilon_0} \tilde{P}(r,t) \, .
\end{eqnarray}
Next, we write $\tilde{E}(r,t) = \sqrt{V} \alpha(t) \chi(r)$ with $\chi(r)$ denoting a single longitudinal mode function normalized according to $\int d^3r |\chi(r)|^2 = 1$ and satisfying the Helmholtz equation $(c^2 \Delta + \omega_c^2) \chi(r) = 0$. Here, $V$ denotes the mode volume defined by the requirement $\sqrt{V}\, \textrm{Max}(|\chi(r)|)] =1$. For a Gaussian mode with radius $w_0$ in a standing wave resonator of length $L$ one calculates $V = \frac{1}{4} \pi w_0^2 L$.
Using the approximation $\frac{1}{2}(\omega_p-\frac{\omega_c^2}{\omega_p})$ $= \frac{1}{2\omega_p}(\omega_p+\omega_c)(\omega_p-\omega_c)$ $\approx (\omega_p-\omega_c) \equiv \delta$ one obtains 
\begin{eqnarray}
\label{eq:MaxwellBloch2}
\dot{\alpha}(t) &+& (\kappa + i \delta)\,\alpha(t) = \sigma(t)
\\ \nonumber
\sigma(t) &\equiv& i \frac{\omega_p}{2 \varepsilon_0 \sqrt{V}} \int d^3r \tilde{P}(r,t)\,\chi^{\ast}(r)
\\ \nonumber
\alpha(t) &\equiv& \frac{1}{\sqrt{V}} \int d^3r \tilde{E}(r,t)\,\chi^{\ast}(r)\, .
\end{eqnarray}
We consider a sample of $N$ two-level atoms with dipole operator $D$, ground state $|g\rangle$ and excited state $|e\rangle$, and assume for the moment that all atoms reside at the same position $r_a$, however, not necessarily where the coupling is maximal, such that each atom is described by the same density matrix $\rho(r,t) \equiv \rho(r_a,t)\,\delta(r-r_a)$. The real polarization then reads $P(r,t) = N\, \textrm{Tr}[\rho(r,t) D])$ $= N \delta(r-r_a) (\rho_{ge}(r_a,t) D_{eg}  + \rho_{eg}(r_a,t) D_{ge})$ with $D_{ge} \equiv \langle g| D |e\rangle$ and $\rho_{ge} \equiv \langle g| \rho |e\rangle$. Using the components of the Bloch vector $u(t) \equiv \rho_{eg}(r_a,t) + \rho_{ge}(r_a,t)$ and $v(t) \equiv i (\rho_{eg}(r_a,t) - \rho_{ge}(r_a,t))$ and $f_a \equiv \sqrt{V} \chi(r_a)$ one gets $\sigma(t) = \frac{\sqrt{2}\,\omega_p D_{eg}}{4 \varepsilon_0 V} N f(r_a) (u(t )+iv(t))$. Introducing the electric field per photon as $\alpha_0 \equiv \sqrt{\frac{\hbar \omega_p}{\varepsilon_0 V}}$, the Rabi-frequency per photon $g \equiv \frac{\sqrt{2}\,D_{eg} \alpha_0}{\hbar}$, the scaled electric field $\beta(t) \equiv \alpha(t)/\alpha_0$, and $U(t) \equiv N u(t)$, $V(t) \equiv N v(t)$ leads to
\begin{eqnarray}
\label{eq:MaxwellBloch2}
\dot{\beta}(t) &+& (\kappa + i \delta)\,\beta(t) = i \frac{g}{4} f_a^{\ast} (U(t) + i V(t))) .
\end{eqnarray}

\begin{figure}
\includegraphics[scale=0.45, angle=0, origin=c]{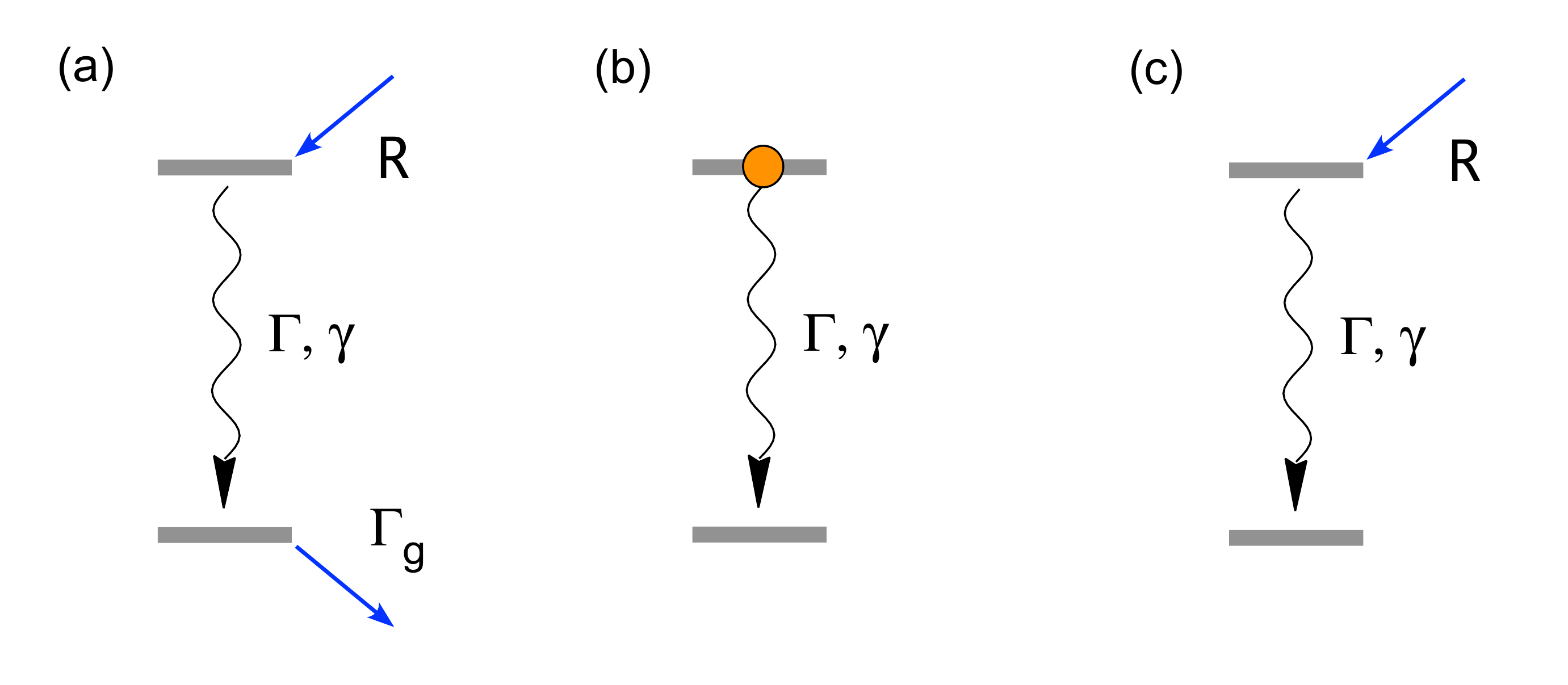}
\caption{\label{Fig.S1} (a) Laser scheme with the upper laser level pumped at the rate $R$ while depumping of the lower laser level at a rate $\Gamma_g$. $\Gamma$ and $\gamma$ denote the decay rate of the upper laser level and of the atomic polarization, respectively. (b) Pulsed laser operation with $R = \Gamma_g = 0$ and $N$ atoms initially populating the upper laser level. (c) Pulsed laser operation with $\Gamma_g = 0$ and the upper laser level pumped at a rate $R(t)$.}
\end{figure}

With regard to the atomic dynamics it is assumed that all atomic dipole moments in perfect synchonization respond to the same cavity mode such that they sum up to form a macroscopic dipole, which follows the optical Bloch equation written in the co-rotating frame after applying the rotating wave approximation.
\begin{eqnarray}
\label{eq:MaxwellBloch3}
\frac{\partial}{\partial t} \left( \begin{array}{cccc} U  \\ V \\ W \\ N \end{array} \right) = 
\left( \begin{array}{cccc} -\gamma & -\delta & 0 & 0  
\\ \delta & -\gamma & -\omega_1 & 0 
\\ 0 & \omega_1 & -(\Gamma+\frac{1}{2}\Gamma_g) & -(\Gamma-\frac{1}{2}\Gamma_g)
\\ 0 & 0 & \frac{1}{2}\Gamma_g & -\frac{1}{2}\Gamma_g \end{array} \right) 
\left( \begin{array}{cccc} U  \\ V \\ W \\ N \end{array} \right) + \left( \begin{array}{ccc} 0  \\ 0 \\ R \\ R \end{array} \right) 
\nonumber \\
\end{eqnarray}
with $W(t) \equiv (\rho_{ee}-\rho_{gg}) N$ denoting the inversion and $\gamma$ and $\Gamma$ denoting the decay rates of the atomic dipole and the atomic inversion, respectively \cite{Coh:92}. Furthermore, $R$ and $\Gamma_g$ denote the rates with which the upper laser state $|e\rangle$ is pumped and the lower laser state $|g\rangle$ is depumped (cf. Fig.~\ref{Fig.S1}(a)). The Rabi-frequency is $\omega_1 \equiv \frac{\sqrt{2}}{\hbar}\, \tilde{E}(r_a,t)\, D_{eg}$ $= \frac{\sqrt{2}}{\hbar} \alpha_0 \,D_{eg} \,f_a\, \beta(t)$ $= g\,f_a \,\beta(t)$. In the remainder of these notes, the co-rotating basis $|g\rangle$, $|e\rangle$ is chosen to rotate with the Bohr frequency of the atomic transition, i.e., $\delta = 0$, which immediately entails $U=0$. Its phase is chosen such that $\beta$ is real and $\omega_1$ is positive. Furthermore, we account for a position spread $r_{\nu}, \nu = 1,...N$ of the atoms inside the laser mode by replacing $|f_{a}|^2 $ by the bunching factor $B \equiv V N^{-1} \sum_{\nu=1}^N | \chi(r_{\nu})|^2$.

\subsection{Pulsed operation after instantaneous inversion}
In this section, pulsed laser operation is considered. The preparation of complete inversion is assumed to be instantaneously achieved at time $t=0$. In a subsequent section, we extend the description to include a more realistic pumping process with a finite duration. Let us then assume $N$ atoms prepared in $|e\rangle$ at time $t=0$ with $R = \Gamma_g = 0$ (cf. Fig.~\ref{Fig.S1}(b)), such that $N$ is constant, and (upon neglect of collisions) there is no decoherence except for spontaneous decay leading to $\Gamma = 2 \gamma$. Furthermore, we focus on the "bad cavity" limit and thus may integrate out the light field. This, by means of Eqs.~(\ref{eq:MaxwellBloch2}) and (\ref{eq:MaxwellBloch3}) leads to the set of equations
\begin{eqnarray}
\label{eq:las1}
\beta^2 &=& \frac{g^2 B}{16 \kappa^2} \,V^2 
\\ \label{eq:las2}
\dot{V} &=& \left(\frac{g^2 B}{4 \kappa} W - \gamma \right) \,V
\\ \label{eq:las3}
\dot{W} &=& - \frac{g^2 B}{4 \kappa} V^2 - \Gamma (W + N)\, ,
\end{eqnarray}
Note that Eq.~(\ref{eq:las2}) sets a threshold for an exponentially growing atomic polarization $V$ given by $W_{\textrm{th}} = \frac{4}{B C} =  \frac{1}{B \eta}$ with the cooperativity defined as $C \equiv \frac{g^2}{\kappa \gamma}$ and $\eta \equiv \frac{24 F}{\pi k_p^2 w_0^2}$ denoting the Purcell factor. The latter merely depends on the cavity finesse $F$, the waist $w_0$ and the wave number $k_p = \omega_p/c$ of the Gaussian laser mode and quantifies the ratio of scattering into the cavity mode as compared to all other vacuum radiation modes \cite{Pur:46, Tan:11}. With our definitions of $\alpha_0$, $g$, and $C$ and the mode volume $V = \frac{1}{4}\pi w_0^2 L$ ($L =$ length of cavity mode) we have $C = 4 \eta$. For times $t>0$ and $t\ll \min\{\gamma^{-1}, \Gamma^{-1}\}$, such that in Eq.~(\ref{eq:las2}) and Eq.~(\ref{eq:las3}) $\gamma$ and $\Gamma$ can be neglected, respectively, a simple class of solution exists
\begin{eqnarray}
\label{eq:sec1}
\beta^2 &=& \frac{\eta B \Gamma}{8 \kappa} N^2 \textrm{sech}^2( \frac{1}{2} N \eta B \Gamma\,(t-t_d) ) 
\\ \label{eq:sec2}
V &=& N \textrm{sech}( \frac{1}{2} N \eta B \Gamma \,(t-t_d) )
\\ \label{eq:sec3}
W &=& -N \tanh( \frac{1}{2} N \eta B \Gamma \,(t-t_d) ) \, ,
\end{eqnarray}
which describes the emission of an intense light pulse far shorter than the natural decay time $1/\Gamma$, scaling with the square of the atom number and thus indicating its collective nature. Note that the number of initially excited atoms $N$ equals the number of photons emitted in the secant pulse of Eq.~(\ref{eq:sec1}), i.e., $2 \kappa \int_{-\infty}^{\infty}d\tau\, \beta^2(\tau) = N$.

\subsection{Pulse width, mean delay time, delay time distribution}
We are now in the position to calculate the width $\Delta t$ (FWHM) as well as the mean delay time $\bar{t}_d$ of the light pulse of Eq.~(\ref{eq:sec1}). First, by setting $\frac{1}{2} = \textrm{sech}^2(\frac{1}{2} N \eta B \Gamma \,(t-t_d)) 
$ one finds 
\begin{eqnarray}
\label{eq:width}
\Delta t = \frac{4 \ln(1 + \sqrt{2})}{N \eta B \Gamma} \, .
\end{eqnarray}
In order to evaluate $\bar{t}_d$ we first determine the mean intra-cavity photon number during a short time-period $0\leq t<\kappa^{-1}$. Since $\kappa^{-1} \ll \Gamma^{-1}$, during this period basically no atom has undergone spontaneous decay yet and hence the average rate of spontaneous photons emitted into the cavity is given by $\bar{R}_0 = N \eta B \Gamma$. Accounting for the photon loss rate $2\kappa$ of the cavity, one finds the initial mean photon number resulting from spontaneous emission of photons into the cavity $\bar{n}_0 \equiv \frac{\bar{R}_0}{2\kappa} = N \eta B \frac{\Gamma}{2\kappa}$. By setting $\bar{n}_0 = \beta(t=0)^2$ with $\beta(t)$ taken from Eq.~(\ref{eq:sec1}) and solving with respect to $t_d$ one finds
\begin{eqnarray}
\label{eq:delay}
\bar{t}_d  = \frac{\ln(N)}{N \eta B \Gamma} \, .
\end{eqnarray}

In order to model the shot-to-shot fluctuations of the delay time observed in our experiment, we calculate the delay time probability. For short times $0\leq t \ll \bar{t}_d$ the probability of a photon being emitted into the cavity by a single atom is $p \equiv (1- e^{-\Gamma t})\eta B$ $\approx \eta B \Gamma t$. Hence, for $N$ atoms the probability that $n_0$ photons have been emitted into the cavity at time $t$ is given by a binomial distribution $P_N(n_0,p) \equiv [N ! /(n_0 !(N-n_0) !)] \,p^{n_0} (1-p)^{N-n_0}$. For $n_0 \ll 1$ a good approximation is $P_N(n_0,p) \approx \exp(-n_0 \ln(\frac{1}{pN})) \exp(-pN)$, i.e., we obtain an exponential dependence on the photon number $n_0$. Hence, we assume that the initial photon number in the cavity at short times on the order of a few cavity lifetimes $\kappa^{-1}$ is associated with the probability distribution $P_{i}(n_0) \equiv \bar{n}_0^{-1}\exp(-n_0/\bar{n}_0)$ with the mean photon number $\bar{n}_0$. By setting $n_0 = \beta(t=0)^2$ with $\beta(t)$ taken from Eq.~(\ref{eq:sec1}) and calculating $dn_0 = - 2 \kappa \,\bar{n}_0 n_0 \,dt $, one finds $P_{i}(n_0)\,dn_0 = P_{d}(t_d) \,d t_d$ with the delay time probability distribution 
\begin{eqnarray}
\label{eq:delay_probab}
P_{d}(t_d) \equiv N^2 \eta B \Gamma \, e^{-N \eta B \Gamma t_d} \exp(- N \, e^{-N \eta B \Gamma t_d}) 
\end{eqnarray}
Remarkably, this relation for $\eta=1$ and $B=1$, which corresponds to the case, when in absence of a cavity a homogeneous sample of atoms radiates into all vacuum modes, reproduces an analytical expression reported in Ref.~\cite{Gro:82} (Eq.~5.36) based upon a quantum mechanical description of the system in terms of Dicke states. A numerical account of delay time statistics, agreeing well with this analytical expression, is discussed in Ref.~\cite{Haa:81}.

\begin{figure}
\includegraphics[scale=0.55, angle=0, origin=c]{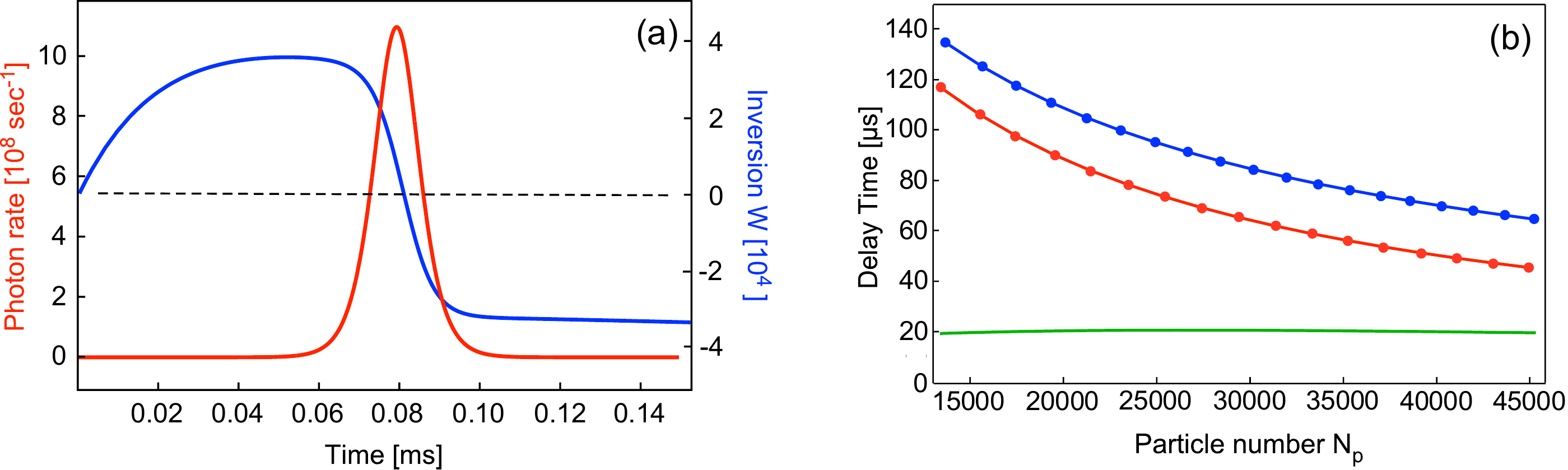}
\caption{\label{Fig.S2} (a) The red and blue line graphs show the photon rate $2 \kappa \,\beta^2(t)$ and the inversion $W(t)$, respectively. (b) The blue and red graphs show the delay time for $\tau_p =21\,\mu$s and $\tau_p = 0$, respectively. The green graph shows their difference.}
\end{figure}

\subsection{Including pump dynamics}
We now extend the analysis to include the dynamics associated with the preparation of inversion during a short period after $t=0$. We find that for preparation periods shorter than $\bar{t}_d$, given in Eq.~(\ref{eq:delay}), merely a constant delay offset adds to $\bar{t}_d$, which is independent of $N$, such that the analytic analysis of the previous section remains valid. As illustrated in Fig.~\ref{Fig.S1}(c), we describe the loading of the upper laser level $|e\rangle$ by the rate $R(t) \equiv  \frac{N_0}{\tau_p} e^{-t/\tau_p}$, where $N_0$ is the total number of atoms pumped into $|e\rangle$. No loss is assumed for the population of $|g\rangle$, i.e., $\Gamma_g = 0$. The incoherent loading contributes to the decay of coherence according to $2 \gamma = \Gamma + R/N_0$. The non-zero rate $R(t)$ is associated with the number of atoms pumped to the system $N(t) = \int_{0}^{t}ds R(s) = N_0 (1-e^{-t/\tau_p})$. A consideration analogous to that before Eq.~(\ref{eq:delay}) leads to the mean number of initial spontaneous intra-cavity photons $\bar{n}_{0}(t) = \frac{\eta B \Gamma}{2\kappa} N(t)$ and an associated rate $\frac{d}{dt}\bar{n}_{0} = \frac{\eta B \Gamma}{2\kappa} R$. The laser equation for the field amplitude Eq.~(\ref{eq:MaxwellBloch2}) can be rewritten as $\frac{d}{dt}\beta^2 = 2 \beta \frac{d}{dt} \beta =  -2\kappa\, \beta^2 -  \frac{1}{2} g \sqrt{B}  \, \beta \, V + \frac{d}{dt}\bar{n}_{0}$ with the extra source term $\frac{d}{dt}\bar{n}_{0}$ added, to account for the initial incoherent scattering of spontaneous photons into the cavity during the pump phase, which requires that $\tau_p$ is sufficiently short such that no coherent dipole moment $V$ can build up during the formation of inversion. Under this condition, the laser equations Eqs.~(\ref{eq:MaxwellBloch2}),(\ref{eq:MaxwellBloch3}) read
\begin{eqnarray}
\label{eq:plas1}
\dot{\beta}(t) &=& -\kappa\, \beta(t) -  \frac{1}{4} g \sqrt{B}  \, V(t) + \frac{\frac{d}{dt}\bar{n}_{0}(t)}{2 \beta(t)}
\\ \label{eq:plas2}
\dot{V}(t) &=&  - g \sqrt{B}\, \beta(t) \,W(t) - \gamma \, V(t)
\\ \label{eq:plas3}
\dot{W}(t) &=& g \sqrt{B}\, \beta(t) \, V(t) - \Gamma \, (W(t) + N(t)) + R(t) \, .
\end{eqnarray}
These equations are numerically solved with the initial conditions $\beta(0)=V(0)=W(0)=0$ using the parameters $g = \sqrt{2\eta \kappa \Gamma}$ and $\eta$, $\kappa$, $\Gamma$ as specified in the main text. A typical example with $N_0=4.8 \times 10^4$ and $\tau_p = 21\, \mu$sec is illustrated in Fig.~\ref{Fig.S2}(a) with the red and blue line graphs showing the intra-cavity photon number $\beta^2(t)$ and the inversion $W(t)$, respectively. In contrast to the case of instantaneous pumping, the number of particles $N_p \equiv 2 \kappa \int_{0}^{\infty}d\tau\, \beta^2(\tau)$ actually contributing a photon to the superradiant pulsed emission is only a fraction of the total number $N_0$ of particles pumped into the upper laser level. For example, for the pulse shown in Fig.~\ref{Fig.S2}(a) with $N_0=4.8 \times 10^4$, one calculates $N_p =3.4 \times 10^4$ corresponding to the red trace of experimental data in Fig.2 of the main text. The delay time and the photon rate is slightly overestimated as compared to the experimental observations, which we attribute to the circumstance that for $\tau_p = 21\, \mu$sec, a coherent dipole begins to form before the formation of inversion is completed, i.e., the approximation at the basis of Eqs.~(\ref{eq:plas1}) is only moderately well fulfilled. In Fig.~\ref{Fig.S2}(b), the blue and red graphs show the delay times versus $N_p$ calculated for exponential pumping with $\tau_p =21\,\mu$s, which corresponds to the situation in the experiments, and for instantaneous pumping ($\tau_p =0$), as considered in the previous section. The green graph shows their difference, which turns out to be independent of the particle number $N_p$ in good approximation. The main message of this analysis is that a pump phase with finite duration can be modeled by an instantaneous formation of inversion, if a particle number independent offset is added to the pulse delay times.

\begin{figure}
\includegraphics[scale=0.45, angle=0, origin=c]{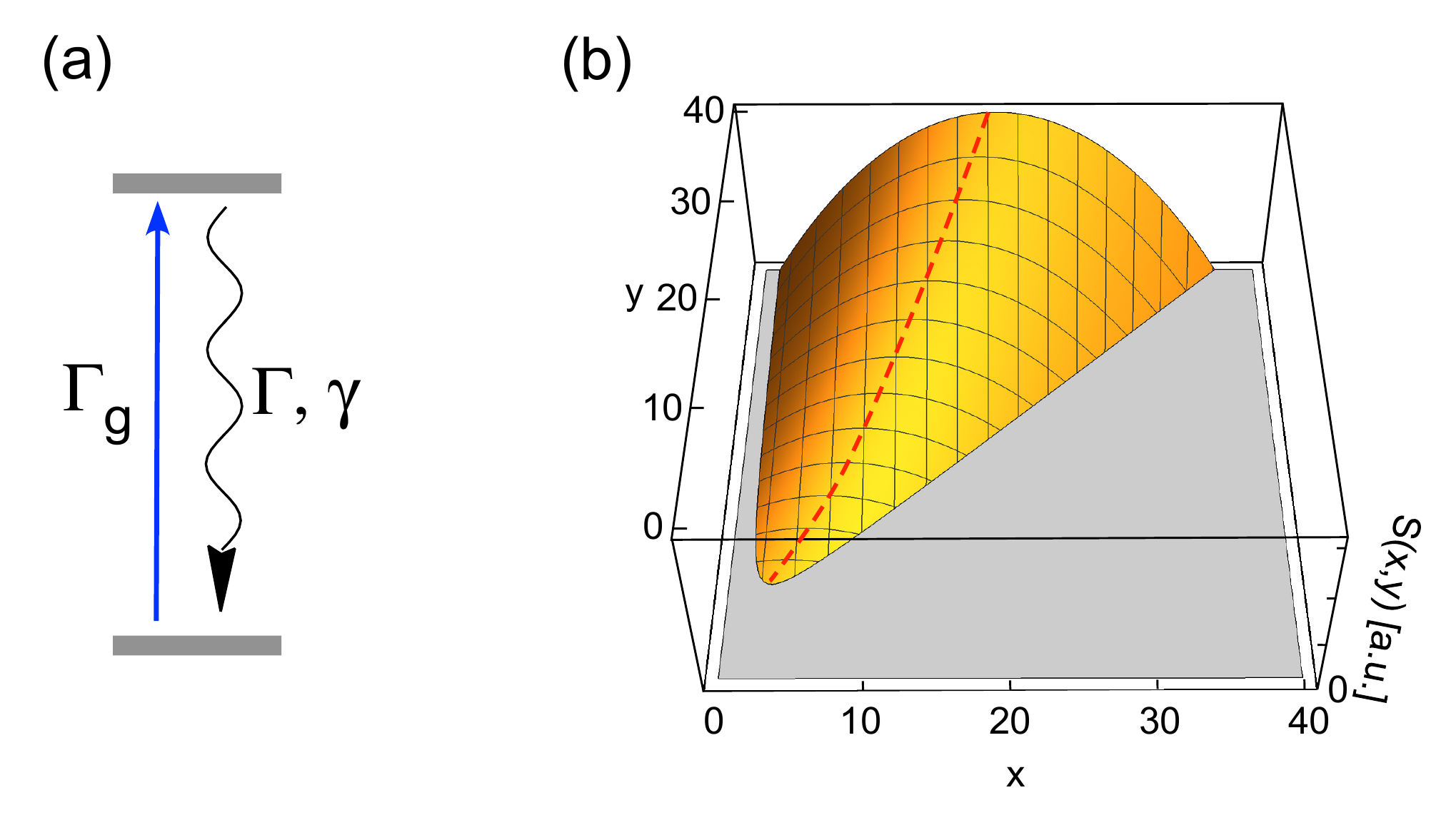}
\caption{\label{Fig.S3} (a) Continuous wave operation by repumping the lower to the upper laser level at a rate $\Gamma_g$. (b) The rate of photons $S(x,y)$ emitted from the cavity is plotted versus the pump strength parameter $x \equiv \Gamma_g/\Gamma$ and the scaled atom number $y \equiv \eta B N$. The dashed red line highlights $S$ for optimal pump strength $x_{\textrm{max}}$.}
\end{figure}

\subsection{Continuous mode operation}
Henceforth, we consider the case, when the population of the lower laser level is continuously repumped to the excited laser level, i.e., $R = N \rho_{gg} \Gamma_g$ (cf. Fig.~\ref{Fig.S3}(a)). Note that the repumping of $|e\rangle$ introduces excess decoherence of the atomic polarization in addition to that by spontaneous decay, i.e., $2\gamma = \Gamma + \Gamma_g$. As in the preparations to Eqs.~(\ref{eq:las1})-(\ref{eq:las3}), $\delta = 0$ is chosen such that $U=0$, and we replace $|f_{a}|^2 $ by the bunching factor $B$. Starting with Eq.~(\ref{eq:MaxwellBloch3}) then yields $\frac{\partial}{\partial t} N(t) = 0$ and hence constant $N$ together with the dynamical equations 
\begin{eqnarray}
\label{eq:las4}
\dot{\beta} &=& -\kappa\, \beta -  \frac{1}{4} g \sqrt{B}  \, V
\\ \label{eq:las5}
\dot{V} &=&  - g \sqrt{B}\, \beta \,W - \gamma \, V
\\ \label{eq:las6}
\dot{W} &=& g \sqrt{B}\, \beta \, V - (\Gamma + \Gamma_g) \, W -  (\Gamma-\Gamma_g) \, N  \, .
\end{eqnarray}
These equations have a stationary solution. Setting $\dot{V} = 0$ in Eq.~(\ref{eq:las5}), one gets $\bar{V} =-g \sqrt{B} \, \bar{W} \beta / \gamma$ and, upon inserting this result into Eq.~(\ref{eq:las4}), one finds $(\bar{W}-\frac{1+x}{\eta B}) \beta = 0$ with the repumping parameter $x \equiv \Gamma_g/\Gamma$. Hence, above threshold, i.e. if $\beta > 0$, one obtains $\bar{W}(\beta > 0) = \frac{1+x}{\eta B}$, while below threshold ($\beta= 0$) Eq.~(\ref{eq:las6}) leads to $\bar{W}(\beta = 0) = \frac{x -1}{x +1}\, N$. At threshold  $\bar{W}(\beta > 0) = \bar{W}(\beta = 0)$ and hence the particle number required at threshold is $N = \frac{1}{\eta B} \frac{(x+1)^2}{(x - 1)}$. By directly inserting $\bar{V}$ and $\bar{W}(\beta > 0)$ into Eq.~(\ref{eq:las6}) one obtains the relation for the rate of photons emitted from the cavity in terms of the scaled particle number $y \equiv \eta B N$ and the repumping parameter $x$: 
\begin{eqnarray}
\label{eq:las7}
S(x,y) \equiv \beta^2 2 \kappa = \frac{\Gamma}{2 \eta B} [y(x-1)-(x+1)^2] \, . 
\end{eqnarray}
In Fig.~\ref{Fig.S3}(b) $S(x,y)$ is plotted, showing that for a given particle number $N$ there is a minimal pump rate, leading to laser emission and a maximal pump rate beyond which laser emission is terminated again, due to excessive decoherence. An analogous plot based upon a full quantum treatment is found in Ref.~\cite{Mei:09}. One may calculate the optimal pumping rate $x_{\textrm{max}}$ where $S$ attains a local maximum with respect to $x$. The result is $x_{\textrm{max}}=\frac{1}{2} y -1$ and $S(x_{\textrm{max}},y) = \frac{\Gamma}{\eta B} (\frac{1}{8} y -1) y$. Likewise, one finds the simple analytic expression $y = (x+1)^2 / (x-1)$ for the threshold in the $xy$-plane by setting $S(x,y) = 0$. In our experimental system, the value of $y$ is approximately $100$ and, hence, the optimal pump strength is $x_{\textrm{max}} \approx 50$ and $S(x_{\textrm{max}},y) \approx 10^{9}$. With $\Gamma = 2 \pi \, 375\,$Hz this would require $\Gamma_g \approx 2 \pi \,18.7\,$kHz.

\subsection{Technical considerations, data analysis, terminology}
\textit{Superradiant laser cavity (SLC):} Precisely controlled tuning of the SLC resonance with respect to the $^{1}$S$_{0} \rightarrow ^{3}$P$_{1}$ transition frequency at $657\,$nm is achieved by actively stabilizing the SLC to a diode laser beam at $780.2\,$nm locked to a Doppler-free resonance of rubidium atoms (with a servo bandwidth of 100 kHz) and sent through an electro-optic fiber modulator (EOFM) tunable between $400$ and $1000\,$MHz. Adjusting the EOFM driving frequency, the SLC resonance at $657\,$nm can be tuned with a precision of a few Hz. The stability of the detuning $\delta_{\textrm{ac}}$ between the resonances of the SLC and the atomic transition is controlled by repeatedly monitoring the transmission an ultra-stable reference laser at $657\,$nm through the cavity (cf. next section). The fluctuations of $\delta_{\textrm{ac}}$ are below a few hundred kHz on all relevant time scales, i.e., much less than the cavity linewidth of $2.3\,$MHz.

\textit{Reference laser (RL) at $657\,$nm:} This RL is frequency stabilized to better than ten Hz to a reference cavity with a finesse of $10^5$ and mirrors optically contacted to a $10\,$cm ultralow expansion glass spacer, which is temperature stabilized to better than a millikelvin and hence exhibits a frequency drift with respect to the stabilized SLC undetectable on the scale of the width of its transmission resonances ($\approx 2.3\,$MHz) during days \cite{Sch:01}. The RL is only used to find the resonance frequency of the calcium atoms such that the SLC resonance at $657\,$nm can be tuned to coincide with the atomic resonance. To achieve this, the RL emission first is tuned to the center of a fluourescence spectrum of the $^{1}$S$_{0} \rightarrow ^{3}$P$_{1}$ transition, Doppler broadened to $20\,$MHz, which is obtained from an atomic beam. Secondly, the SLC resonance at $657\,$nm is tuned in resonance with the RL. Fine tuning is obtained by adjusting the SLC until maximal superradiant emission is observed. Finally, the RL is tuned to be maximally transmited through the SCL. This adjustment is repeatedly controlled during measurements.   

\textit{Photon detection:}
The detector for the superradiant emission is a standard photon counting module (COUNT-20C) purchased from Laser Components Inc. with $73\, \%$ quantum efficiency and 20 dark counts per second and $45\,$ns dead time. The detector for the incoherent emission is a standard photomultiplier tube (PMT H10720, Hamamatsu). Background noise is no problem for the detection of superradiant emission. Detector saturation is prevented by using calibrated attenuators. Due to a small spatial observation angle, for detection of the incoherent emission, averaging over hundreds of traces was necessary for decent signal to noise. 

\begin{figure}
\includegraphics[scale=0.45, angle=0, origin=c]{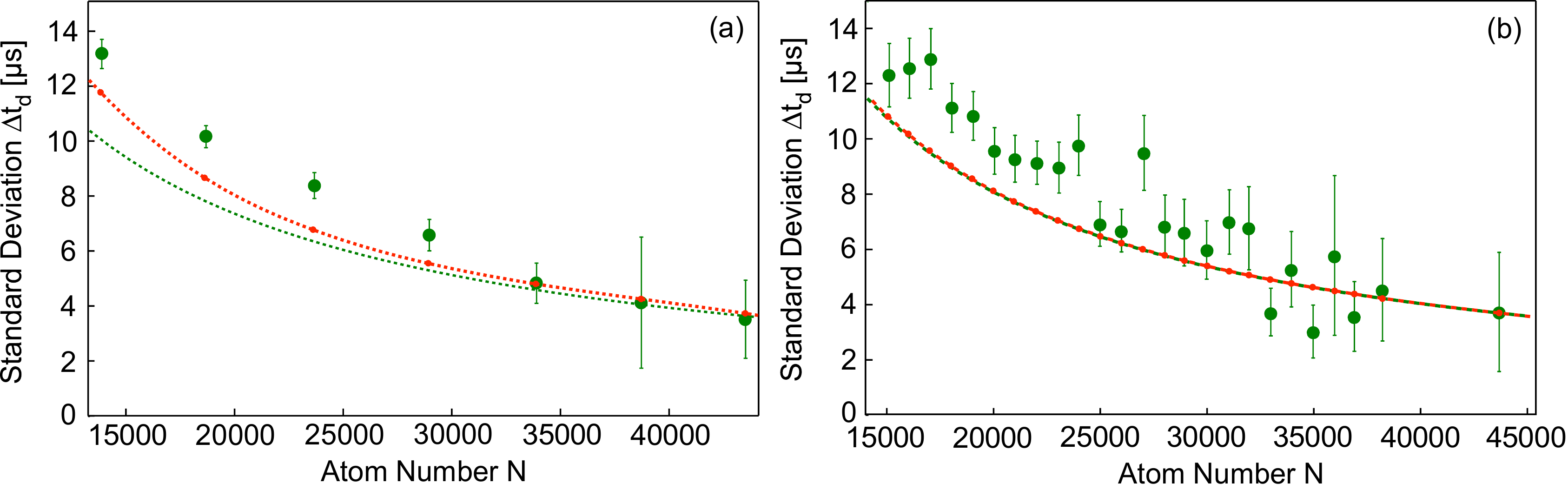}
\caption{\label{Fig.S4} (a) The green disks show the standard deviations found for all pulses belonging to the same atom number class with error bars showing the corresponding standard deviations of the mean. The dashed green line shows the standard deviation of the pulse delay time as calculated from the Gumbel distribution in Eq.~(\ref{eq:delay_probab}). The small red disks show the expected standard deviations including a systematic contribution due to the finite partition used for the atom number axis. The dashed red line is an interpolation for eye guiding. In (b) the same data set is analysed with a fivefold finer partition of the $N$-axis.}
\end{figure}

\textit{Data analysis:}
As pointed out in the main text, the number of atoms $N$ contributing to each observed pulse is determined by fitting with a hyperbolic secant model using the same fixed values of the bunching parameter $B$ and the Purcell factor $\eta$. We thus obtain a data set of pulses with atom numbers $N_{i,\nu} \in [N_i - \delta N_i, N_i + \delta N_i]$ falling within intervals $\mathcal{N}_i = [ N_i - \delta N_i, N_i + \delta N_i ]$ of equal width $2\delta N_i = N_{i+1}-N_i$, which form a regular partition of the $N$-axis. In Fig.3(c) of the main text, which is repeated here in Fig.~\ref{Fig.S4}(a), we have chosen a relatively coarse partition, in order to provide large numbers of pulses in each atom number class and hence better statistics at the cost of a small systematic overestimation of the plotted standard deviations of the pulse delay time for small $N$.  The green disks in Fig.~\ref{Fig.S4}(a) show the standard deviations with respect to all pulses belonging to the same atom number class. The error bars show the corresponding standard deviations of the mean. The green dashed line graph shows the theoretically expected standard deviation versus the atom number according to the Gumbel distribution in Eq.~(\ref{eq:delay_probab}). The small red disks shows a more refined calculation of the expected standard deviations accounting for the finite partition used for the $N$-axis. The plotted quantity is $\sigma_i \equiv \sqrt{\sigma_{1,i}^2+\sigma_{2,i}^2}$, where $\sigma_{1,i}$ denotes the average of the standard deviations of all Gumbel distributions in Eq.~(\ref{eq:delay_probab}) evaluated at atom numbers $N_{i,\nu} \in \mathcal{N}_i $, and $\sigma_{2,i}$ is the standard deviation of all mean delay times $\bar{t}_d(N_{i,\nu})$ evaluated according to Eq.~(\ref{eq:delay}) with $N_{i,\nu} \in \mathcal{N}_i$. The second contribution $\sigma_{2,i}$ accounts for the fact that even in absence of quantum fluctuations quantified by $\sigma_{1,i}$, the use of a finite $N$-partition gives rise to statistical fluctuations. For comparison, we show in Fig.~\ref{Fig.S4}(b) the same data as in (a), however analyzed using a fivefold finer partition of the $N$-axis. This should give rise to $5 \times 7 = 35$ data points, rather than the shown 25, however, for this partition, some classes $\mathcal{N}_i$ comprise only a single pulse such that no standard deviation can be calculated. For this partition, the partition-dependent contribution to the shown standard deviations is negligible, i.e., the green and red dashed line graphs practically overlap. On the other hand, the fluctuations of the standard deviations (as indicated by the error bars) are significantly larger as compared to (a) due to limited numbers of pulses per atom number class. \linebreak 

\textit{Technical parameter drifts and fluctuations:}   
All technical fluctuations, besides those of the detuning $\delta_{\textrm{ac}}$ between the atomic resonance and the cavity resonance (which is initially adjusted to zero), exclusively translate into fluctuations of the number of atoms contributing to superradiance. This applies for fluctuations of the optical pumping frequencies or intensities as well as the frequencies and intensities of the laser beams of the the magneto-optic traps (MOT) and the Zeeman slower. Since we determine atom numbers ex post by fitting superradiant pulses with a secant model with the bunching parameter $B$ and the Purcell factor $\eta$ fixed, such fluctuations are not a problem. At the shot-to-shot time scale (10 s) the atom number is stable to better than a few percent, as absorption imaging after ballistic expansion shows. On longer time scales above ten minutes, the atom number fluctuates by $\pm 25 \%$. The main source of such drifts is the laser system of the MOT operating on the principle fluorescence line in the singlet system, which serves as a reservoir of pre-cooled atoms for loading the MOT in the triplet system. This laser source consists of a dual mode titanium sapphire standing wave laser, emitting on two longitudinal modes with 1 GHz frequency difference, which are used for sum-frequency mixing in an LBO crystal placed in an enhancement cavity. Due to small thermal drifts of the laser alignment, on a time scale of several minutes a third weakly contributing longitudinal mode can arise, which reduces the power in the two modes relevant for sum-frequency generation, while the overall power remains constant. The resulting atom number drift is passed on to the lattice of $^{3}$P$_{0}$-atoms. A reliable absolute determination of the number of atoms participating in superradiant emission with imaging techniques is not possible since we cannot selectively address the $^{3}$P$_{1}, m=0$ level, and even less just those $^{3}$P$_{1}$ atoms that will contribute to superradiance rather than to spontaneous emission. 

Fluctuations of $\delta_{\textrm{ac}}$  could lead to false particle numbers in the secant fitting procedure and hence have to be carefully prevented. Absolute stability of $\delta_{\textrm{ac}}$ is obtained by locking the cavity to a laser operating near $780\,$nm that is stabilized to a Doppler-free rubidium resonance (with $100\,$kHz servo bandwidth). The obtained stability of $\delta_{\textrm{ac}}$ is controlled by repeatedly monitoring the transmission of our ultra-stable reference laser at $657\,$nm through the cavity. The fluctuations of $\delta_{\textrm{ac}}$ are below a few hundred kHz on all relevant time scales, which should be compared with the cavity linewidth of $2.3\,$MHz. Hence, fluctuations of $\delta_{\textrm{ac}}$ should not contribute significant errors in the atom number determination with the secant model. The limited significance of $\delta_{\textrm{ac}}$ fluctuations is also reflected by the circumstance that the secant model works exceptionally well for practically all pulses observed, with the same values for the fixed parameters $B$ and $\eta$, which is not expected, if these pulses corresponded to differences in $\delta_{\textrm{ac}}$ rather than different atom numbers. 

\textit{Sources of decoherence:}
In contrast to Ref.~\cite{Nor:16a}, in our experiments, decoherence via atom loss with the result of a threshold, is not a significant issue, due to the short duration of the superradiant emission on the $100\,\mu$s time-scale. Another possible source of decoherence is inhomogeneous line broadening. The most prominent source of inhomogeneous broadening of the laser transition is a deviation from parallel alignment of the linear polarization of the magic lattice and the magnetic bias field (5 Gauss), which could degrade the magic of the lattice. For $20\,$MHz lattice depth, we estimate that inhomogeneous broadening should be below $\Gamma_{\textrm{inh}} = 2 \pi \,100\,$kHz. To roughly estimate a tentative threshold, one may equate: $\Gamma_{\textrm{col}} = \Gamma_{\textrm{inh}}$ with the collectively enhanced emission rate $\Gamma_{\textrm{col}} \equiv N C \Gamma$ (C= cooperativity, $\Gamma$ = natural linewidth, $N =$ atom number), i.e., $N =  \Gamma_{\textrm{inh}} / (\Gamma C)$, which evaluates to $1.3 \times 10^4$ for our system \cite{Nor:16b}. In fact, we never observe pulses with particle numbers below $10^4$.

\textit{Superradiance versus lasing:}
We use the term "superradiant laser" for a laser operating in the bad cavity regime $\tau_c \ll \tau_e$, where $\tau_c$ denotes the cavity lifetime and $\tau_e$ the lifetime of the upper laser state. Coherence is predominantly stored in the atomic polarization, however, the cavity takes the role to shape the spatial mode of the emitted light to form a directed beam. The light propagation time through the inverted medium $\tau_d \equiv d/c$, where $d$ is the diameter of the inverted medium and $c$ the speed of light, trivially satisfies  $\tau_d \ll \tau_c$, such that even if the cavity is removed, the inverted medium can emit cooperatively. This is the case of superradiance in the Dicke sense. If the cavity operates in the good cavity regime $\tau_e \ll \tau_c$, one may call this conventional lasing. Here, it is predominantly the cavity that acts as the flywheel that stores coherence in addition to shaping the spatial mode of the emitted light. This regime, however, does not exclude the possibility that the inverted medium upon removal of the cavity emits cooperatively. Note that this entire conceptualization does not depend on how the inversion is pumped, continuously or pulsed.

\end{document}